\documentclass{mn2e}
\usepackage{epsf,times}
\newcommand{\etal}{{et al}\/.}

\begin{document}
\title[{\rm Chandra} observation of 3C\,296]{A {\it Chandra} observation of the X-ray environment and jet of 3C\,296}
\author[M.J.~Hardcastle \etal]{M.J.\ Hardcastle$^{1,2}$, D.M.\ Worrall$^2$,
M.\ Birkinshaw$^2$, R.A.\ Laing$^{3,4}$ and A.H.\ Bridle$^5$\\
$^1$ School of Physics, Astronomy and Mathematics, University of
Hertfordshire, College Lane, Hatfield, Hertfordshire AL10 9AB\\
$^2$ Department of Physics, University of Bristol, Tyndall Avenue,
Bristol BS8 1TL\\
$^3$ European Southern Observatory, Karl-Schwarzschild-Strasse 2,
D-85748 Garching-bei-M\"unchen, Germany\\
$^4$ University of Oxford, Department of Astrophysics, Denys Wilkinson
Building, Keble Road, Oxford OX1 3RH\\
$^5$ National Radio Astronomy Observatory, 520 Edgemont Road,
Charlottesville, VA 22903-2475, U.S.A\\
}
\maketitle
\begin{abstract}
We have observed the twin-jet radio galaxy 3C\,296 with {\it Chandra}.
X-ray emission is detected from the nucleus, from the inner parts of
the radio jet, and from a small-scale thermal environment around the
jet deceleration region. As we have found in previous observations of
other twin-jet radio galaxies, the X-ray jet and a steep pressure
gradient in the external thermal environment are associated with the
region where strong bulk deceleration of the jet material is suggested
by radio observations. Our observations provide additional evidence
that the inner jets of twin-jet objects are always associated with
a relatively cool, dense central X-ray emitting component with a short
cooling time.
\end{abstract}

\section{Introduction}

In an earlier paper (Hardcastle \etal\ 2002) we reported on {\it
Chandra} X-ray observations of the twin-jet FRI radio galaxy 3C\,31.
It has been accepted for some time that the jets in this class of
radio galaxy often have relativistic bulk speeds on parsec scales but
trans-sonic or sub-sonic speeds on scales of hundreds of kiloparsecs,
so that they must decelerate on intermediate scales. The available
observational evidence from studies of large samples and of individual
objects (e.g.\ Laing \etal\ 1999; Laing \& Bridle 2002a) suggests that
this deceleration happens on scales of $\sim 1$--10 kpc for FRIs of
intermediate to high power, corresponding to scales of a few
arcseconds at the 100-Mpc distances of typical well-studied objects.
It is thus possible to study the deceleration region with {\it
Chandra} and the NRAO Very Large Array (VLA).

Combined X-ray and radio observations of the jet deceleration regions
are important for two reasons. Firstly, all models in which the jet
decelerates by mass loading require an external pressure gradient to
prevent the jet from decollimating and disrupting. The hot, X-ray
emitting phase of the ISM/IGM is the only one with sufficient pressure
to be dynamically important in this situation. By observing the X-ray
emitting medium on scales of a few kpc we can test the jet
deceleration model both qualitatively (the steep pressure gradient
must be detected for standard models to operate) and quantitatively
(by comparing with the observations predictions of the external
pressure based on detailed modelling of the radio emission). Secondly,
the internal dissipation of energy that arises from bulk deceleration
can help to explain the X-ray synchrotron jets seen in the inner few
kpc of many twin-jet FRIs (e.g. Worrall \etal\ 2001, 2003; Hardcastle
\etal\ 2001). The X-ray jet emission may itself be an indication of
bulk deceleration, and there is some direct evidence for this in the
dynamics of the nearest such jet, Cen A (Hardcastle \etal\ 2003).

In our {\it Chandra} observations of 3C\,31 we detected both an X-ray
synchrotron jet and a kpc-scale thermal environment with a steep
pressure gradient. Subsequently it was possible to show (Laing \&
Bridle 2002b) that the form of the external pressure profile was
similar to that predicted from detailed modelling of the radio jet
properties (Laing \& Bridle 2002a), and to derive good constraints on
the kinetic luminosity of the jet. The brightest X-ray emission occurs
in the region of the jet immediately before the rapid deceleration
inferred from radio modelling. These observations of 3C\,31 provided
strong support for the jet deceleration model, but it is important to
explore whether this model applies to other objects. We report here on
the result of our {\it Chandra} observations of 3C\,296, which is
another twin-jet radio galaxy suitable for modelling in this way.

3C\,296 is a well-known FRI radio source (Birkinshaw, Laing \& Peacock
1981; Leahy \& Perley 1991; Hardcastle \etal\ 1997) hosted by the
elliptical galaxy NGC 5532 at a redshift of 0.0237, a similar redshift
to that of NGC 383 which contains 3C\,31. Like NGC 383, NGC 5532 lies
in a group of galaxies, including at least one bright nearby galaxy,
NGC 5531; the association of the group galaxies has been confirmed
spectroscopically (Miller \etal\ 2002). In the X-ray, 3C\,296 was
detected with {\it Einstein} (Fabbiano \etal\ 1984), the {\it ROSAT}
HRI (Hardcastle \& Worrall 1999) and the {\it ROSAT} PSPC (Miller
\etal\ 1999). Radio sidedness studies (Hardcastle \etal\ 1997) suggest
that the jet decelerates significantly on scales of about 10 arcsec,
making it a good candidate for {\it Chandra} observations.

Throughout the paper we use a concordance cosmology with $H_0 = 70$ km
s$^{-1}$ Mpc$^{-1}$, $\Omega_{\rm m} = 0.3$ and $\Omega_\Lambda =
0.7$. At the redshift of 3C\,296, 1 arcsec corresponds to 480 pc.
Spectral indices $\alpha$ are the energy indices and are defined in
the sense $S_{\nu} \propto \nu^{-\alpha}$. The photon index $\Gamma$
is $1+\alpha$.

\section{The observations}

3C\,296 was observed for 49429 s with the {\it Chandra} ACIS-S on 2003
August 31. The source was positioned near the standard aim point on
the back-illuminated S3 CCD. We reprocessed the data in the standard way
to apply the latest calibration files (using CIAO 3.1 and CALDB 2.28).
As the data were taken in VFAINT mode, we used `VFAINT cleaning' to
identify and reject some background events. We also removed the
0.5-pixel event position randomization and removed streaking from the
S4 chip.

\begin{figure*}
\epsfxsize 17cm
\epsfbox{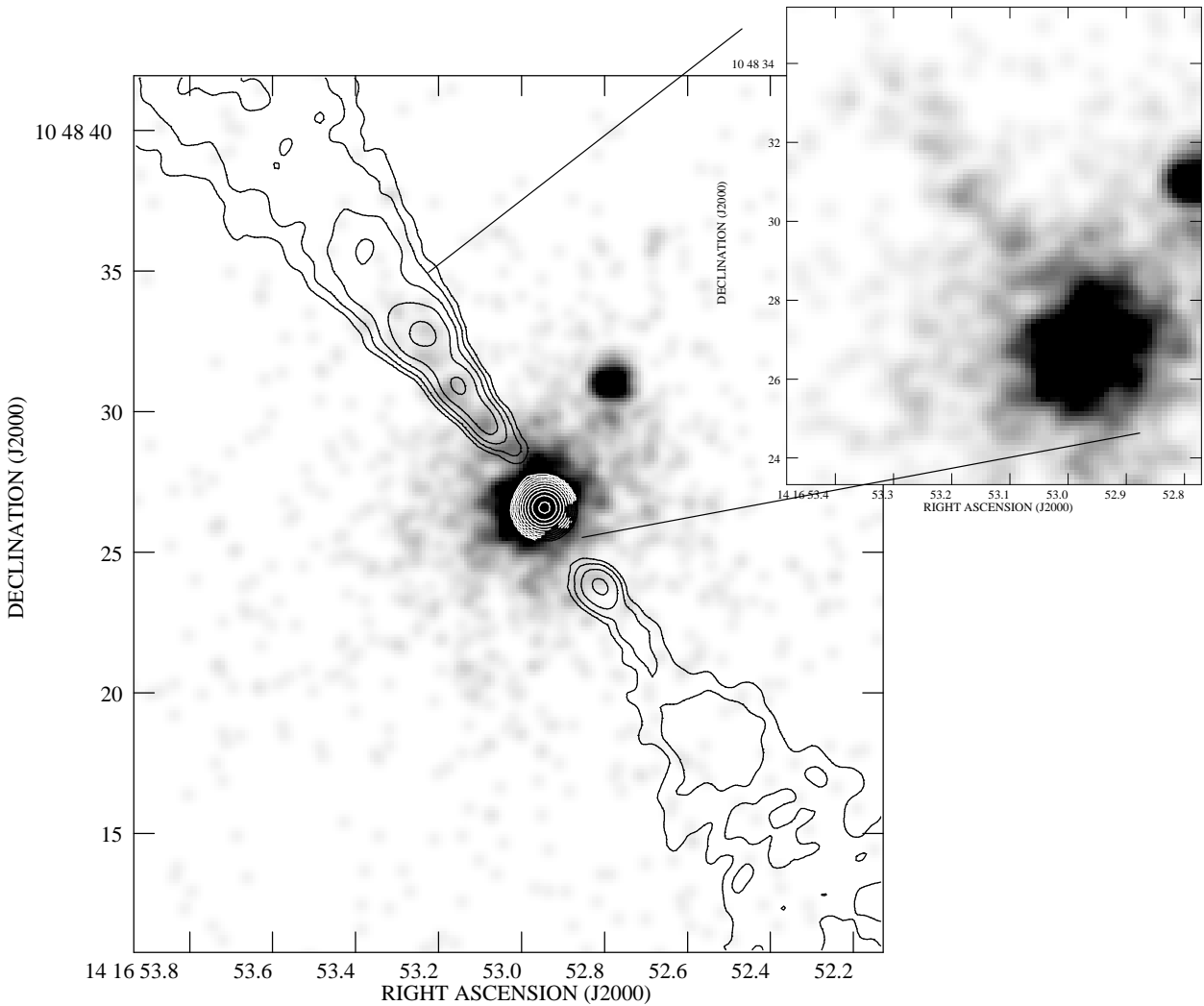}
\caption{{\it Chandra} image of the inner regions of 3C\,296. The
greyscale is the X-ray data binned in 0.123-arcsec pixels and smoothed
with an 0.5-arcsec (FWHM) Gaussian: black is 1 count pixel$^{-1}$.
Superposed are contours from the 8.4-GHz, 0.75-arcsec resolution radio
map obtained with the VLA by Hardcastle \etal\ (1997); the contours
are drawn at $100 \times (1, 2, 4, \dots)$
$\mu$Jy beam$^{-1}$. The X-ray point source to the NE is discussed in
section \ref{pointsource}. The inset shows the inner part of the jet
without radio contours, but with the same greyscale level.}
\label{image}
\end{figure*}

We used events in the energy range 0.5--5 keV to construct an image
(throughout the paper this energy range is used for imaging). As Fig.
\ref{image} shows, there is a clear detection of an X-ray core and
jet, as well as extended emission extending out to scales of tens of
arcsec. In addition, on larger scales, X-ray emission is detected from
the nearby galaxy seen in {\it HST} images (Martel \etal\ 1999 class
it as an S0) and from NGC 5531. The spatial and spectral properties of
these components are discussed in the following sections. Spectra were
extracted and corresponding response matrices constructed using {\sc
ciao}, and model fitting was carried out (in the energy range 0.4-7.0
keV) in {\sc xspec}. In all cases, except where otherwise stated, the
extracted spectra were grouped so that each fitting bin had $>20$
counts after background subtraction. Errors quoted throughout are the
$1\sigma$ value for one interesting parameter, unless otherwise
stated.

\section{The X-ray core}
\label{core}

As we will show in Section \ref{gas}, the thermal emission from the
host galaxy of 3C\,296 has a small angular scale. It is therefore not
possible spatially to isolate the X-ray nucleus completely. We
extracted a spectrum from an circle of radius 1.5 arcsec, taking
background from an annulus between 1.5 and 2.5 arcsec. These small
regions were intended to minimize the contribution from thermal
emission and also avoid the nearby point source and resolved jet. There are 914
net counts in the 0.4--7.0 keV energy range in this region.
Unsurprisingly, the resulting spectrum is not well fitted ($\chi^2 =
66.3$ for 38 degrees of freedom) by a single power-law model with
Galactic absorption ($N_{\rm H} = 1.85 \times 10^{20}$ cm$^{-2}$) and
shows large residuals at around 1 keV. The fit is unimproved if the
absorbing column is allowed to vary. A combination of a power law and
a {\sc mekal} model with 0.5 solar abundance, both with Galactic absorption,
is a better fit ($\chi^2 = 41.4$ for 36 d.o.f.) but gives an unusually flat
photon index, $\Gamma = 0.77_{-0.08}^{+0.09}$; the temperature of the thermal
component in this model is $0.61 \pm 0.07$ keV. The best fits ($\chi^2
= 27.6$ for 35 d.o.f.) are obtained if we allow the power-law model to
be intrinsically absorbed: in this case the intrinsic column is $(1.0
\pm 0.5) \times 10^{22}$ cm$^{-2}$, the photon index is $1.6 \pm 0.4$
(the errors for these two quantities are $1\sigma$ for two interesting
parameters, as they are correlated), and the {\sc mekal} model has $kT =
0.62 \pm 0.05$ keV. In this model, almost all the counts below 1 keV
are contributed by the thermal component, and this interpretation is
supported by imaging, which shows the source to be much more
point-like above 1 keV than below it. The power-law component would
contribute $530_{-160}^{+240}$ counts to our extraction region in this
model, or roughly half of the total, and its normalization corresponds
to an unabsorbed 1-keV flux density of 24 nJy; its absorbed 2--10 keV flux is
$1.7 \times 10^{-13}$ ergs cm$^{-2}$ s$^{-1}$. No additional component
(e.g. a second unabsorbed power law) gives a significant improvement
over this fit.

Intrinsic obscuration with large columns is rare, but not unknown, in
FRI sources (e.g. Hardcastle \etal\ 2002; Donato, Sambruna \& Gliozzi
2004; Evans \etal\ in prep.). The inferred column density for 3C\,296
is among the largest yet reported for an FRI, comparable to that seen
in 3C\,270 (Gliozzi, Sambruna \& Brandt 2003). It is noteworthy that
the two FRI sources known to have high X-ray absorption columns,
3C\,270 and 3C\,296, also both show an unresolved nuclear component in
the optical but not the UV (Chiaberge \etal\ 2002); this is possible
evidence that the optical and X-ray emission in these sources
originate on the same spatial scales.

Although large for an FRI source, the column density seen in 3C\,296 is
much less than the columns of a few $\times 10^{23}$ cm$^{-2}$
inferred for obscuring `tori' in FRI sources like Centaurus A (e.g.\ Evans
\etal\ 2004) or FRIIs like Cygnus A (Ueno \etal\ 1994). As we have pointed out
previously (e.g. Hardcastle \etal\ 2002) the presence or absence of a
moderate absorbing column (which could be related to the known
presence of cold, dusty gas in the centres of these galaxies) in the
nuclear X-ray emission does not provide any direct information on the
presence or absence of an obscuring torus around the central AGN if
the X-ray emission is jet-dominated and originates on scales larger
than that of the AGN; in this scenario we would see evidence for a
torus only in sources with very weak jets (Cen A) or very strong AGN
(Cyg A). We will return to this point elsewhere (Evans \etal\ in
prep.).

Small-scale thermal emission is common in fits to cores of these
systems, and we will discuss this component further in Section
\ref{gas}. The abundance of the thermal component is not well
constrained if we allow it to be a free parameter, since the absorption
column and normalization of the power-law component of the fit can be
adjusted to allow for any value of the ratio between the line emission
and continuum emission in the {\sc mekal} model. Our choice to fix the
value to 0.5 solar is consistent with Section \ref{gas}.

\section{The X-ray jet}
\label{jet}

X-ray emission is detected from the northern jet, starting at around 2
arcsec from the nucleus and extending to about 10 arcsec. The spatial
scale corresponds closely to the bright inner region of the radio jet
(Fig.\ \ref{image}), immediately before the location where the radio
sidedness data imply a sudden deceleration (Hardcastle \etal\ 1997).
We extracted a spectrum in a rectangular region 7.1 arcsec long and
2.6 arcsec wide, aligned with the jet and starting at 2 arcsec from
the nucleus. There are 74 net counts in the 0.4--7.0 keV energy range,
after subtracting background from an identical region at a different
position angle with respect to the nucleus (this background
subtraction procedure was chosen to ensure subtraction of the correct
level of background due to the extended thermal emission discussed in
Section \ref{gas}). We binned the extracted spectrum with 18
counts per spectral bin to avoid discarding data. The spectrum is well
fitted ($\chi^2 = 1.4$ for 2 d.o.f.) with a single power law with
Galactic absorption: the spectral index $\alpha$ is $1.0 \pm 0.4$ and
the normalization corresponds to an unabsorbed 1-keV flux density of
$1.2 \pm 0.3$ nJy. This makes the jet substantially fainter (both in
terms of X-ray flux density and X-ray luminosity) than those seen in
some other 3CRR FRI sources, though the X-ray spectral index is
similar.

\begin{figure*}
\epsfxsize 16cm
\epsfbox{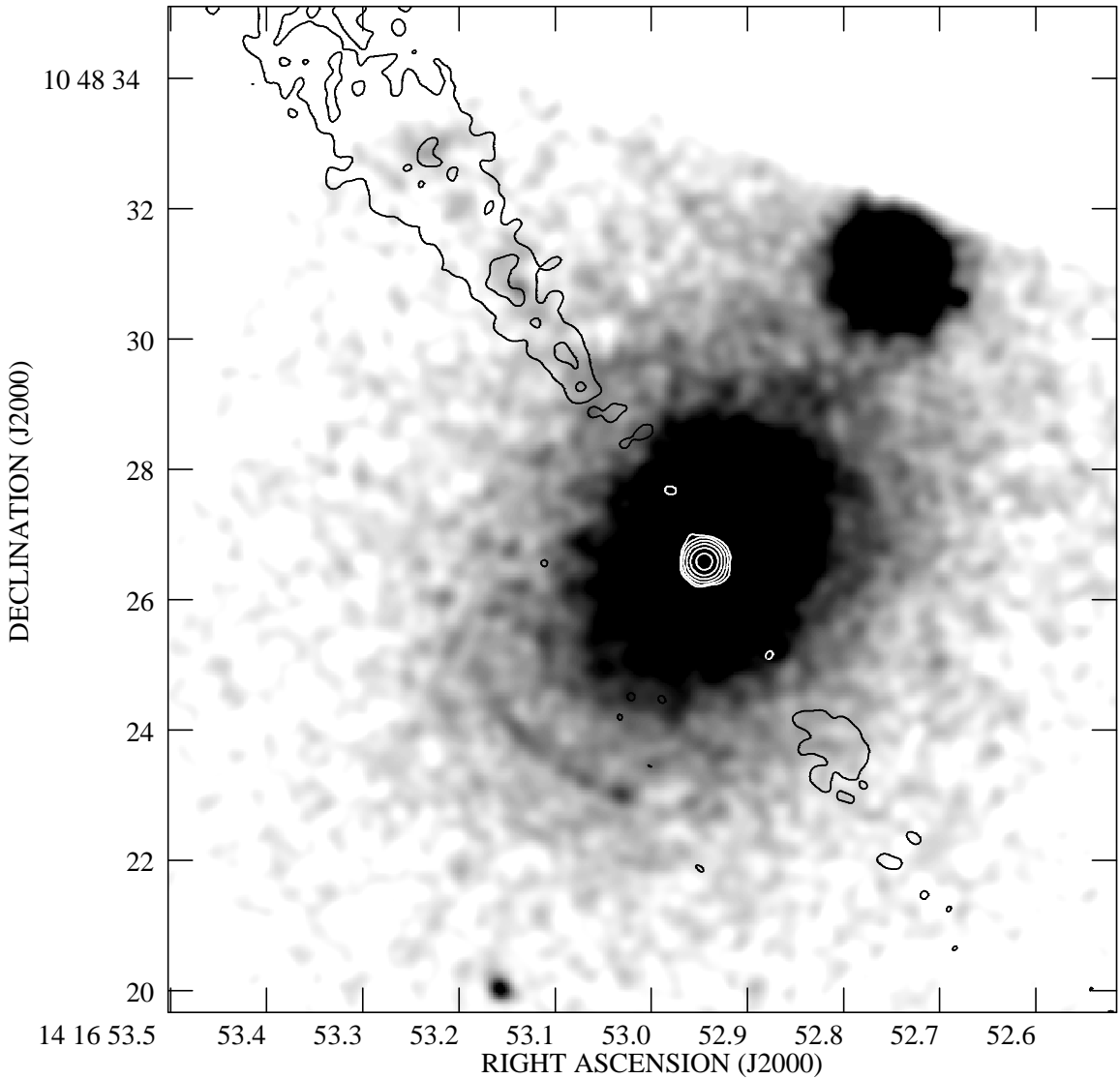}
\caption{The {\it HST} STIS image of 3C\,296 at 2200 \AA, convolved
  with a 0.25-arcsec (FWHM) circular Gaussian. Overlaid are contours
  of the 0.24-arcsec resolution 8.4-GHz radio map of Hardcastle \etal\
  (1997) at $60 \times (1,4,16...)$ $\mu$Jy beam$^{-1}$. There is a
  weak but significant excess of emission in the position angle of the
  jet. The bright source to the NW is the optical counterpart of the
  X-ray point source in Fig.\ \ref{image}, and is discussed in Section
  \ref{pointsource}. The arc-like feature SW of
  the core is discussed in Section \ref{lensing}.}
\label{stisjet}
\end{figure*}

\begin{figure}
\epsfxsize \linewidth
\epsfbox{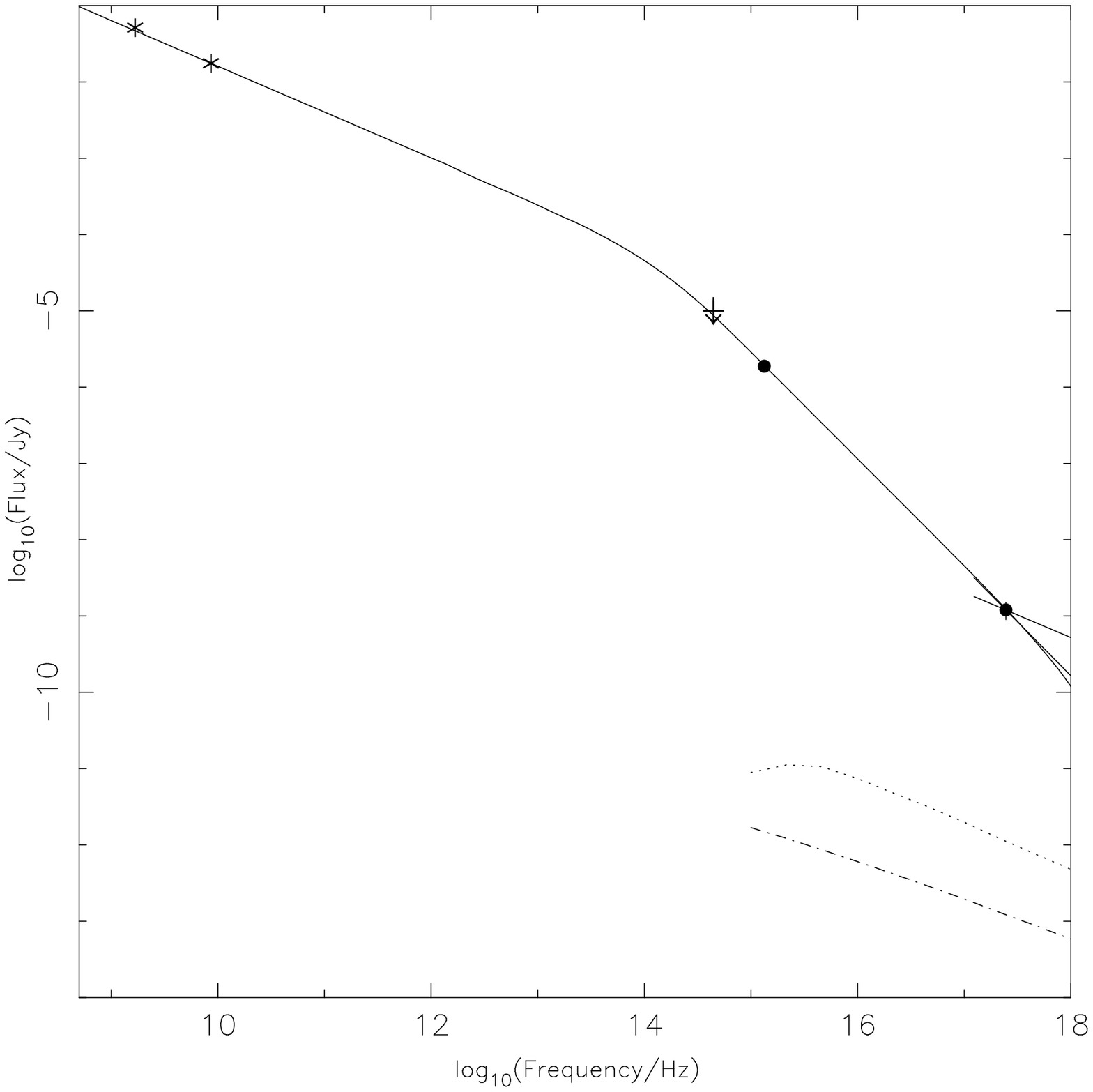}
\caption{The radio-to-X-ray spectrum of the jet of 3C\,296, based on
  the region detected with {\it Chandra}. Radio points (stars) are
  taken from the data of Hardcastle \etal\ (1997); the optical data
  points are from archival {\it HST} data, as described in the text.
  The `bow tie' around the X-ray flux density illustrates the
  $1\sigma$ range of spectral index in the power-law fit to the jet
  spectrum. The solid line shows a broken power-law synchrotron model,
  with low-frequency spectral index 0.55 and high-frequency spectral
  index 1.4. The dotted and dash-dotted lines show the predicted X-ray
  emission for inverse-Compton scattering of the cosmic microwave
  background and for the synchrotron self-Compton process
  respectively, assuming minimum-energy conditions.}
\label{jetspectrum}
\end{figure}

The 8.4-GHz flux density in the same region is 17.5 mJy, so
that the two-point radio-to-X-ray spectral index is $0.96 \pm 0.01$,
somewhat steeper than is seen in other 3CRR sources (the median value
for known jets is 0.92), but comparable to the extended emission in
Cen A (Hardcastle \etal\ 2003). Since the radio spectral index in the
inner jets is 0.6 (Hardcastle \etal\ 1997) fitting the now standard
synchrotron model to the radio and X-ray data points requires a break
in the synchrotron spectrum; locating the break requires constraints
from optical observations. There is no evidence for an optical jet in
the {\it Hubble Space Telescope} Wide Field/Planetary Camera 2 image
taken in the F702W filter (Martel \etal\ 1999) after subtraction of a
model of the elliptical isophotes: based on the level of structure in
the residual image, we place a limit on the jet flux density at $4.34
\times 10^{14}$ Hz of $\la 10$ $\mu$Jy. Similarly, we were unable to
detect jet emission in ground-based optical images. However, there is
a weak jet detection in the {\it HST} STIS image centred around 2200
\AA\ and taken in 2000 (Allen \etal\ 2002), extending out to the edge
of the STIS field of view, 8.5 arcsec from the core. An image of this
jet detection is shown in Fig.\ \ref{stisjet}. After background
subtraction, correction for the relative sizes of the STIS and {\it
Chandra} extraction regions and scaling for an estimated 0.25 mag of
Galactic extinction at 2200 \AA, we estimate a flux density at $1.3
\times 10^{15}$ Hz for the jet region of 1.9 $\mu$Jy. This allows us
to construct a spectrum for the jet that is similar to those of
other sources (Fig.\ \ref{jetspectrum}).

As with other jets of this kind, the inverse-Compton emission
mechanism (assuming no beaming and only synchrotron and CMB photons as
the parent photon population) underpredicts the observed X-ray by
several orders of magnitude (the predicted spectra are plotted on Fig.\
\ref{jetspectrum}) unless there is a large departure from the
equipartition/minimum-energy conditions. We conclude that the jet is
another example of the association between X-ray synchrotron radiation
and the inner, dissipative jet of an FRI source. There is some
evidence that the X-ray-to-radio flux density ratio falls with
distance along the jet, as seen in other jets: the ratio of 1-keV to
8.4-GHz flux densities between 2 and 4 arcsec from the core is $(0.14
\pm 0.05) \times 10^{-6}$, while between 4 and 10 arcsec it is $(0.05
\pm 0.01) \times 10^{-6}$.

No X-ray emission is detected from the counterjet of 3C\,296.
The radio flux density in the equivalent region of the counterjet is
around 6 mJy, so naively we might expect to see around 1/3 of the jet
counts, or 25 in total, from the counterjet region. If the jet
sidedness is due entirely to beaming, then (as we pointed out in the
case of 3C\,31) the expected jet-counterjet ratio in the X-ray is
greater than that in the radio, as a result of the steeper X-ray
spectral index, and in fact we would expect about 20 counts in total.
As the $1\sigma$ uncertainty in the number of counts in the counterjet
region is around 10, we cannot rule out the possibility that the
counterjet really is emitting X-rays at the predicted level. The same
comment applies to the brightest region of the counterjet, the knot at 3
arcsec from the core, where the upper limits are insufficient to rule
out the possibility that this component is emitting at its expected
level of $\sim 7$ counts. We therefore cannot say with these data
whether the jet and counterjet have identical spectra over the radio
to X-ray range.

\section{Nearby and associated objects}
\label{nearby}

\subsection{The nearby point source}
\label{pointsource}

\begin{figure*}
\epsfysize 20cm
\epsfbox{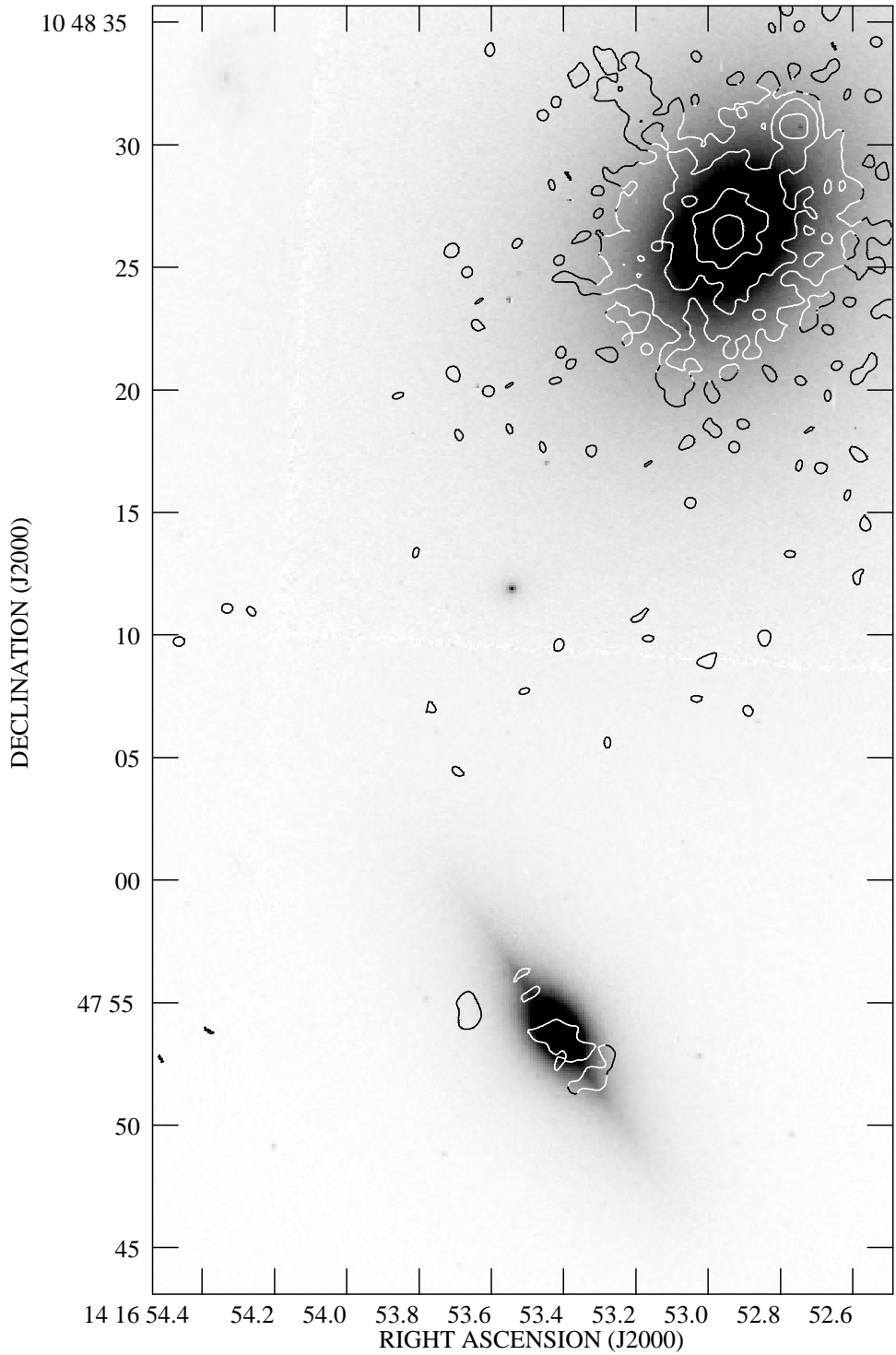}
\caption{The X-ray emission detected by {\it Chandra} superposed on an
  archival {\it Hubble Space Telescope} WFPC2 image of the area around NGC
  5532. The X-ray contours are from the smoothed X-ray image shown in
  Fig.\ \ref{image}; the lowest contour is at approximately $3\sigma$
  above background (Hardcastle 2002). The {\it HST} image is the
  snapshot image in the F702W filter presented by Martel \etal\
  (1999). The X-ray point source to the northwest of 3C\,296 is close
  to a faint optical point source: the extended X-ray source to the
  south is associated with the spiral companion galaxy.}
\label{overlay}
\end{figure*}

The X-ray source 5 arcsec to the NW of 3C\,296's core is associated
with bright UV emission (Fig.\ \ref{stisjet}) and weak optical
emission (Fig.\ \ref{overlay}). In the X-ray its radial profile is
consistent with that of a point source, and its optical counterpart is
also point-like. It contains 135 net {\it Chandra} counts in a
1.2-arcsec source circle, with background taken from a concentric
annulus between 1.2 and 1.7 arcsec, and is well fitted ($\chi^2 = 6.0$
for 4 degrees of freedom) with a power law with Galactic absorption,
but with a remarkably steep spectrum, $\Gamma = 4.4 \pm 0.4$. (A
black-body model with $kT \sim 0.1$ keV is an unacceptably poor fit,
$\chi^2 = 9$ for 4 d.o.f.) Its 1-keV flux density on the power-law
model is 17.8 nJy. Re-examination of the {\it ROSAT} HRI data shows a
weak elongation of the X-rays in the direction of this point source,
consistent with the idea that it was present at that epoch, although
the significance of the elongation is limited in view of the known
problems with {\it ROSAT} aspect reconstruction. There is no evidence
for inter-observation variability in the {\it Chandra} data. The
optical counterpart is only weakly detected in the F702W WFPC2 {\it
HST} image taken in 1994 and shown in Fig.\ \ref{overlay} (flux
density at $4.3 \times 10^{14}$ Hz is 2.3 $\mu$Jy: $R \sim 23$ mag),
but more strongly in an ACS HRC F606W image taken in 2002 (47 $\mu$Jy
at $5.1 \times 10^{14}$ Hz) and is very strong in the STIS image
discussed above (93 $\mu$Jy at $1.3 \times 10^{15}$ Hz). If taken at
face value (neglecting any possible optical variability of the source)
then the object's broad-band spectrum would be strongly peaked in
the UV.

The object's very faint optical magnitude means that it is difficult
to associate it with any Galactic source, particularly given the high
Galactic latitude ($64^\circ$) of 3C\,296. A hot white dwarf in the
Galactic halo would be one possible identification: the X-ray
luminosity is in the observed range, but the extreme optical colour
(or optical variability) argues against such a model. If the X-ray
source were actually at the distance of NGC 5532, its unabsorbed X-ray
flux in the {\it ROSAT} band (0.1--2.4 keV) of $4.8 \times 10^{-13}$
ergs cm$^{-2}$ s$^{-1}$ would give it a luminosity in this band of $6
\times 10^{41}$ ergs s$^{-1}$, which would make it extreme even for an
ultra-luminous X-ray source; in particular, it would be $\sim
1.5$--$2$ orders of magnitude more luminous than ultra-luminous
supersoft sources such as the one in M101 (Mukai \etal\ 2003). We
therefore conclude that it is most likely to be a spectrally peculiar
background AGN, possibly amplified in brightness to some extent by
lensing (see Section \ref{lensing}).

%F702W: 636-666 counts, 280s => 7.1 microJy (1994-12-14).
%F606W: 120 electrons/s (x 2/3) => 47 microJy (2002-07-10)
%STIS: 73.8 microJy => 93.3 after extinction correction. (2000/04/15)

\subsection{The nearby spiral}

As Fig.\ \ref{overlay} shows, the X-ray emission from this galaxy
(which, from its magnitude, is likely to be a member of the NGC 5532
group, although it has not been spectroscopically confirmed to be one)
is mostly associated with the galactic plane, but there is a weak but
significant detection of emission further away from it. There are
insufficient counts from this region to fit a spectrum to the two
components separately. The 60 net counts in the combination of the two
are well fitted ($\chi^2 = 0.60$ for 1 degree of freedom) with a
power-law model with Galactic absorption and a flat spectrum ($\Gamma
= 1.2 \pm 0.3$). The unabsorbed 2--10 keV flux of the galaxy is $1.3
\times 10^{-14}$ ergs cm$^{-2}$ s$^{-1}$, corresponding to a
luminosity of $1.7 \times 10^{40}$ ergs s$^{-1}$ in this band if the
galaxy is at the distance of NGC 5532. This is a plausible
luminosity for a large spiral galaxy.

\subsection{NGC 5531}

Both extended and compact emission is detected from the nearby
elliptical galaxy NGC 5531. In a source circle 30 arcsec in size
(taking background from a concentric annulus between 30 and 40 arcsec)
there are 270 net counts in the 0.4--7.0 keV energy band. Fitting with
a single power-law model with Galactic absorption gives a poor fit
($\chi^2 = 17.4$ for 10 d.o.f.), with residuals below 1 keV, and
fitting a thermal model alone gives an even worse fit; a better fit is
given by fitting a combination of a power-law model and a thermal
({\sc mekal}) component with fixed 0.3 solar abundances ($\chi^2 =
10.8$ for 8 d.o.f.: $kT = 0.44 \pm 0.10$ keV, $\Gamma =
2.2_{-0.7}^{+0.4}$). This, together with the images, suggests that
there is some harder nuclear X-ray emission from NGC 5531, together
with extended thermal emission. The unabsorbed 2--10 keV flux of the
galaxy is $1 \times 10^{-14}$ ergs cm$^{-2}$ s$^{-1}$, almost all
provided by the power-law component, or $1.3 \times 10^{40}$ ergs
s$^{-1}$: given that NGC 5531 is around 1.5 magnitudes fainter than
NGC 5532 in the optical, we cannot rule out the possibility that all
the hard X-ray emission comes from a population of X-ray binaries (see
Section \ref{gas}).

\subsection{The ultraviolet arc SE of NGC 5532.}
\label{lensing}

The ultraviolet feature evident about 2~arcsec (1 kpc) SE of the centre of
NGC 5532 (Fig. 2) has the appearance of a gravitationally-lensed arc.
Such features are now relatively common in clusters of galaxies (for
example in Abell 370; Lynds \& Petrosian 1989), but more usually only
point-like lenses are seen near galaxies (e.g., from the Einstein
cross; Huchra et al. 1985). A simple calculation suggests that NGC
5532 is capable of forming an arc from a background galaxy at a
redshift of about 0.4, if that galaxy lies close to the line of sight
through the nucleus of NGC 5532, and it would be interesting to obtain
an optical spectrum of the brightest region of the arc to check this
interpretation. The arc is not detected in the WFPC2 or ACS {\it HST}
images of the galaxy, presumably due to the higher stellar background
in these images, so that optical observations might be challenging.

The mass distribution capable of lensing an extended object into an
arc could also greatly amplify the brightness of a point object lying
near the same line of sight. This could be an interpretation of the
great over-brightness of the X-rays from the soft source discussed in
Section \ref{pointsource}, provided that this also lies at a
significant redshift. The arc and the soft source could then originate
in the same group of galaxies at redshift $\sim 0.4$.

\section{The extended emission}
\label{gas}

\begin{figure}
\epsfxsize \linewidth
\epsfbox{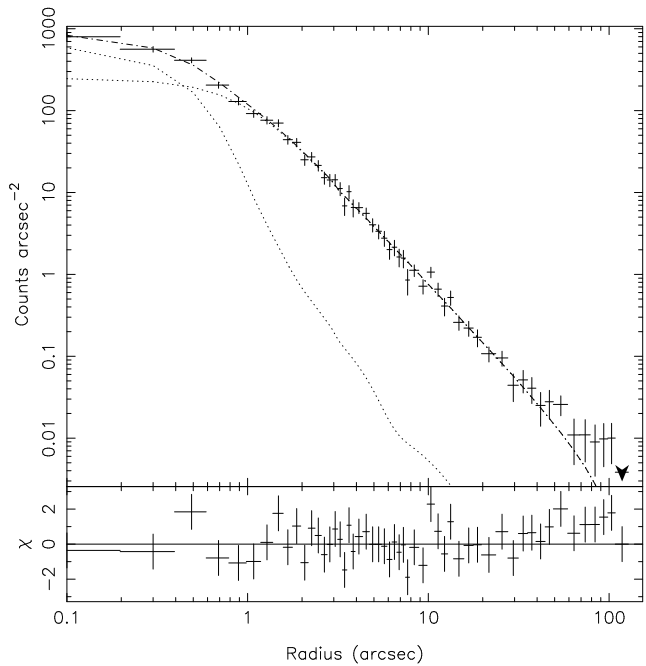}
\caption{The radial profile of the extended emission centred on the
  nucleus of 3C\,296, with the two components of the fitted model
  described in the text, their sum, and the residuals.}
\label{rprofile}
\end{figure}

Spatially resolved emission is seen in the X-ray image of 3C\,296 out
to at least 30 arcsec from the X-ray nucleus, although most of the
counts are on small angular scales. We used the same process as for
3C\,31 (Hardcastle \etal\ 2002) to extract a radial profile from the
0.5--7.0 keV events dataset. The profile was centred on the nucleus,
and excluded all detected point sources, as well as the jet region. It
extended to 110 arcsec from the core, with background being taken
between 110 and 130 arcsec; this ensures that all of the radial
profile region falls on the S3 chip. Exposure correction was made
using a map calculated at the peak energy of the dataset, 0.9 keV. We
fitted the resulting profile (Fig.\ \ref{rprofile}) with a model
consisting of a point source (we use the PSF parametrization of
Worrall \etal\ 2001) and a free $\beta$ model. Unsurprisingly, a
point-source model alone was an unacceptable fit to the data. A good
fit was obtained ($\chi^2 = 48$ for 51 degrees of freedom) by adding a
$\beta$ model with $\beta = 0.56 \pm 0.01$, $\theta_c =
0.7_{-0.10}^{+0.14}$ arcsec (errors are $1\sigma$ for two interesting
parameters). The unresolved core component contained $465 \pm 40$
counts in this model (consistent with the $530^{+240}_{-160}$ counts
in the power-law spectral component), while the central normalization
of the $\beta$ model was $300 \pm 60$ counts arcsec$^{-2}$. There is
little evidence in these fits for extended emission from the 3C\,296
group on larger scales, although the positive residuals at 50--100
arcsec in Fig.\ \ref{rprofile} may give an indication of its presence.
We do not attempt to characterize the large-scale extended emission,
as it clearly makes only a small contribution, if any, to the radial
profile on the S3 chip.

To determine the spectral properties of the extended emission, we used
XSPEC to carry out fits of {\sc mekal} models to annular regions
centred on the galaxy (the regions used are listed in Table
\ref{annuli}), excluding the nearby point source and jet emission
regions. In the central region we included a power-law model with
intrinsic absorption, as discussed in Section \ref{core}. Initially,
we allowed both the temperature and the abundance to vary freely,
while keeping the absorption column fixed at the Galactic value. We
found that this produced poor fits in several annuli, notably the one
between 5 and 12.5 arcsec: there were large residuals at high
energies, though the fits were reasonable below 1 keV. We therefore
fitted each annulus with a combination of a {\sc mekal} model and a
power law with no absorption in excess of Galactic and a fixed photon
index of 1.0, which was intended to represent a contribution from
X-ray binaries. This brought the fitting statistic to an acceptable
level in all cases. The best-fitting abundance was high, with a joint
fit to all four regions giving $Z = 1.4_{-0.4}^{+0.9}$ solar. As we
regard such high metal abundances as a priori unlikely given our
measurements from other sources and other work (e.g. Matsushita,
Ohashi \& Makishima 2000), we have fixed
abundances to 0.5 in the fits tabulated in Table \ref{annuli}. As the
Table shows, the luminosities of the best-fitting power-law components
sum to a few $\times 10^{40}$ ergs s$^{-1}$, which is plausible
for an elliptical galaxy with $M_{\rm B} = -21.8$
(e.g. Matsumoto \etal\ 1997).

\begin{table}
\caption{Regions fitted with thermal models}
\label{annuli}
\begin{tabular}{lrrrr}
\hline
Region&$kT$&Power-law luminosity&$\chi^2$/d.o.f.\\
(arcsec)&(keV)&(2--10 keV) (ergs s$^{-1}$)\\
\hline
0--2.5&$0.68 \pm 0.02$&N/A&71/55\\
2.5--5.0&$0.78_{-0.04}^{+0.04}$&$(1\pm 1) \times 10^{40}$&20/13\\
5.0--12.5&$0.84_{-0.05}^{+0.03}$&$(3_{-1}^{+0.5}) \times 10^{40}$&18/14\\
12.5--30&$0.82 \pm 0.08$&$(1.5 \pm 1) \times 10^{40}$&12/7\\
\hline
\end{tabular}
\vskip 5pt
\begin{minipage}{\linewidth} For the 0--2.5 arcsec region the
  power-law model was fixed to the parameters previously determined
  (Section \ref{core}).
\end{minipage}
\end{table}

\begin{figure*}
\epsfxsize 17.7cm
\epsfbox{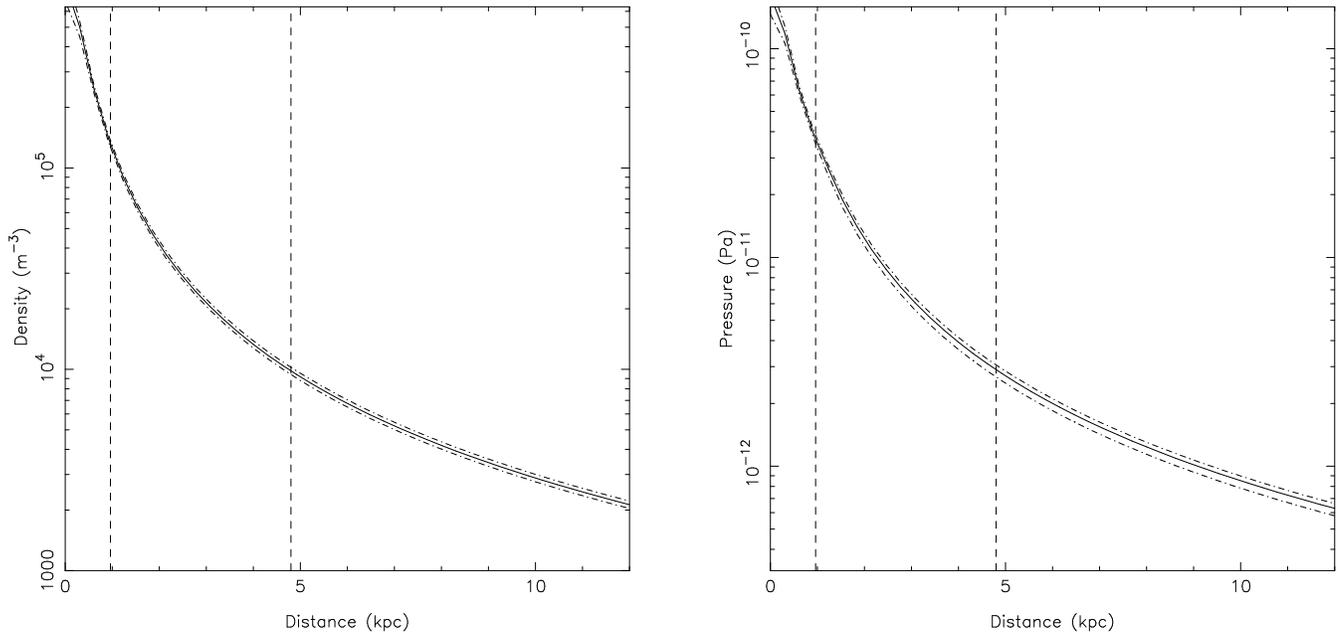}
\caption{Proton density (left) and pressure (right) as a function of
radius in 3C\,296, based on the model described in the text. The solid
black lines show the density and pressure derived from the
best-fitting $\beta$ models ($\beta = 0.56$, $r_c = 0.7$) and the
fitted temperature. The surrounding dotted lines show the combined
$1\sigma$ uncertainties due to the conversion between central
normalization and density, the uncertainties on the $\beta$ model fits
and the uncertainty on temperature. The vertical dotted lines show the
(projected) start and end of the X-ray jet. A linear temperature
gradient between 0.6 and $0.82 \pm 0.04$ keV is assumed in the inner
2.5 arcsec.}
\label{pressure}
\end{figure*}

\begin{figure}
\epsfxsize 8.3cm
\epsfbox{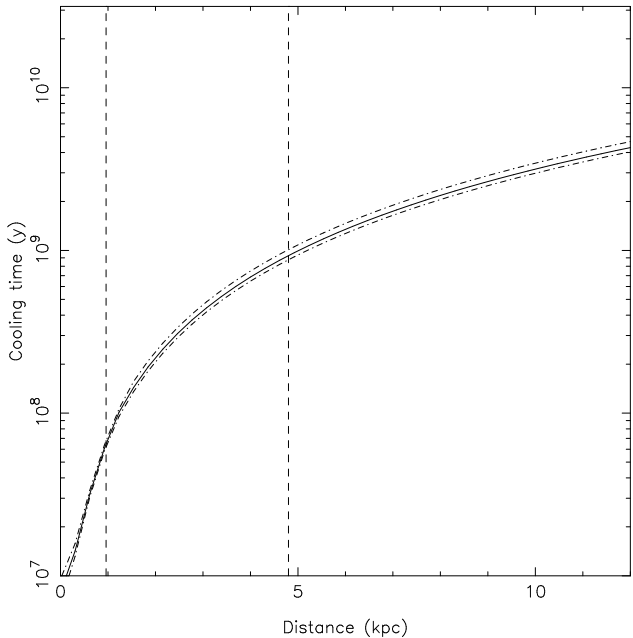}
\caption{Cooling time as a function of radius in 3C\,296. Lines and models as in
Fig. \ref{pressure}.}
\label{cooling}
\end{figure}

Apart from the lower best-fitting temperature in the central few
arcsec, which we also observed in 3C\,31, there is little evidence
here for a temperature gradient: the temperatures beyond 2.5 arcsec
are all consistent at the $1\sigma$ level with a value of 0.82
keV. An onion-skin deprojection similarly provides only marginal
evidence for a temperature gradient. We do not yet know the
temperature of the large-scale group emission, if any, around 3C\,296
(a recent {\it XMM} observation will provide more information when it
becomes available) but in the absence of other data we can model the
atmosphere on the scales of interest as isothermal with a temperature
$kT = 0.82$ keV beyond 2.5 arcsec. Assuming that there is no abundance
gradient, it is then straightforward to determine the external
pressure profile (Birkinshaw \& Worrall 1993). Here we neglect the
hard component in the spectrum which we have associated with X-ray
binaries, since it contributes a small fraction of the count density
on the sky. The central particle density of the $\beta$-model fit is
then $(7 \pm 1) \times 10^5$ m$^{-3}$, and the pressure and density as
functions of radial distance are plotted in Fig.\ \ref{pressure}.
(This assumes an abundance of 0.5 solar; if the abundance were as high
as the best-fitting value, 1.4 solar, the densities and corresponding
pressures would be reduced by a factor 1.5.) As we found for 3C\,31,
there is a steep pressure gradient associated with the region of the
X-ray jet. The pressure gradient here is steeper (${\rm d} \log P/{\rm
d}\log R \sim 1.6$ rather than 1.0 as in 3C\,31) because of the
smaller core radius of the thermal material. The cooling time of the
hot gas that provides the pressure gradient is small, only $\sim 10^7$
years in the centre (Fig. \ref{cooling}), so that (also as with
3C\,31) this component should be transient in the absence of energy
input from the jet or elsewhere. It seems that the association between
a small-scale component of gas with a short cooling time and the jet
may, as we suggested in the case of 3C\,31, be a general requirement
on jet deceleration models, and this again raises interesting questions
about the relationship between the cooling gas and the energy supplied
by the jet.

The minimum pressure in the jet on the assumption of a pure
electron-positron plasma with a minimum electron Lorentz factor of 1,
determined using the spectrum discussed in Section \ref{jet} and
neglecting projection, is less than the external pressure for the
length of the X-ray jet region (as was the case for 3C\,31; Laing \&
Bridle 2002b), by factors of between about 10 and 1.5. This is not
surprising, as there must be additional contributions to the internal
pressure of the jet, and there is no compelling reason to expect the
minimum-energy condition to be satisfied in this region.

\section{Summary and conclusion}

Our {\it Chandra} observations have detected the nucleus and the jet
of 3C\,296 and have resolved the hot gas in the inner parts of the
host galaxy. We have also detected the nearby galaxy NGC 5531, a
bright nearby spiral galaxy, and a soft point source close to NGC
5532.

The non-thermal component of the X-ray core of 3C\,296 has a higher
best-fitting column density for intrinsic obscuration than is typical
in twin-jet FRI sources, but we see neither the high column density
nor the other X-ray features of an obscured active nucleus.

The X-ray jet of 3C\,296 is towards the faint end of the observed
range in X-ray to radio ratio for this type of jet, but the
best-fitting spectral index and the available constraints on the
overall SED of the jet, including a weak detection in the UV with the
STIS instrument on {\it HST}, make us confident that this is another example
of X-ray synchrotron emission from the inner jet of an FRI source. As
with other sources of this kind, the X-ray emission is spatially
closely associated with the region inferred from radio observations to
be where jet deceleration is taking place.

The detection of a dense, small-scale thermal medium in the centre of
3C\,296's host galaxy is in agreement with observations of other FRIs
(Worrall \etal\ 2001, 2003; Hardcastle \etal\ 2001, 2002; Donato
\etal\ 2004) and with the predictions we made based on the 3C\,31
data. A quantitative comparison between the X-ray data and a detailed
model of the radio structure of the jet will be presented elsewhere.

\section*{Acknowledgments}

We are grateful to Dan Evans for helpful discussion of his analysis of
the nuclear X-ray emission of 3C\,296. MJH thanks the Royal Society
for a research fellowship. The National Radio Astronomy Observatory is
a facility of the National Science Foundation operated under
cooperative agreement by Associated Universities, Inc.

\end{document}